# Observation of tightly bound trions in monolayer MoS$_2$

Kin Fai Mak[1], Keliang He[2], Changgu Lee[3], Gwan Hyoung Lee[4], James Hone[4], Tony F. Heinz[1], and Jie Shan[2*]

[1] Departments of Physics and Electrical Engineering, Columbia University, 538 West 120th St., New York, NY 10027, USA
[2] Department of Physics, Case Western Reserve University, 10900 Euclid Avenue, Cleveland, OH 44106, USA
[3] SKKU Advanced Institute of Nanotechnology (SAINT) and Department of Mechanical Engineering, Sungkyunkwan University, Suwon 440-746, Korea
[4] Department of Mechanical Engineering, Columbia University, New York, NY 10027, USA
*Email: jie.shan@case.edu

**Two-dimensional (2D) atomic crystals, such as graphene and transition-metal dichalcogenides, have emerged as a new class of materials with remarkable physical properties[1]. In contrast to graphene, monolayer MoS$_2$ is a non-centrosymmetric material with a direct energy gap[2-5]. Strong photoluminescence[2,3], a current on-off ratio exceeding $10^8$ in field-effect transistors[6], and efficient valley and spin control by optical helicity[7-9] have recently been demonstrated in this material. Here we report the spectroscopic identification in doped monolayer MoS$_2$ of tightly bound negative trions, a quasi-particle composed of two electrons and a hole. These quasi-particles, which can be created with valley and spin polarized holes, have no analogue in other materials. They also possess a large binding energy (~ 20 meV), rendering them significant even at room temperature. Our results open up new avenues both for fundamental studies of many-body interactions and for opto-electronic and valleytronic applications in 2D atomic crystals.**

The trion binding energy that we observe in monolayer MoS$_2$ is nearly an order of magnitude larger than that found in conventional quasi-2D systems, such as semiconductor quantum wells (QWs)[10-13]. This is a consequence of the greatly enhanced Coulomb interactions in monolayer MoS$_2$, arising from reduced dielectric screening in gapped 2D crystals and the relatively heavy carrier band masses associated with the Mo $d$-manifolds[4,5,14]. For an electron density as high as $n = 10^{11}$ cm$^{-2}$, for instance, the dimensionless interaction parameter $r_s$ is ~60 in monolayer MoS$_2$ (Supplementary Information S1). This value is significantly larger than that for carriers in QWs even at very low doping levels[15]. Monolayer MoS$_2$ is a strongly interacting system even in the presence of relatively high carrier densities; it thus presents an ideal laboratory for exploring many-body phenomena, such as carrier multiplication and Wigner crystallization[16].

The atomic structure of MoS$_2$ consists of hexagonal planes of S and Mo atoms in a trigonal prismatic structure (Fig. 1a)[17]. The two sublattices of the hexagonal MoS$_2$ structure are occupied, respectively, by one Mo and two S atoms (Fig. 1b). Monolayer MoS$_2$ is a direct gap semiconductor with energy gaps located at the K and K' points of the Brillouin zone (Fig. 1c). Both the highest valence bands and the lowest conduction bands are formed primarily from the Mo $d$-orbitals[4,17]. The large spin-orbit interaction splits the highest valence bands at the K (K') point by ~ 160 meV[2,3,7,14,18]. The valley and spin (VS) degrees are coupled because of the lack of inversion symmetry in monolayer $MoS_2$. As has been recently shown experimentally, this allows optical pumping of a single valley (and spin) with circularly polarized light[7-9].

Here we investigate the optical response of monolayer $MoS_2$ as a function of carrier density by means of absorption and photoluminescence (PL) spectroscopy. In our investigations we have made use of $MoS_2$ monolayers prepared by mechanical exfoliation. Field-effect transistors (FETs) using $MoS_2$ were fabricated on $SiO_2$/Si substrates; the doping density in the $MoS_2$ channel was systematically varied by applying a voltage to the Si back-gate. All measurements were performed at 10 K, if not otherwise specified. The typical variation of the drain-source current ($I_{ds}$) with the gate voltage ($V_g$) is shown in the inset of Fig. 1d. The response is characteristic of an FET with an *n*-type channel. Spontaneous negative doping, presumably from defects within the $MoS_2$ layer and/or substrate interactions, has been commonly reported in mechanically exfoliated $MoS_2$[6,19]. Over the accessible range of the gate voltage (-100 – +80 V) our $MoS_2$ FETs exhibit electron doping. The low off-state current at high negative gate voltages is understood to reflect electron localization in 2D[15].

The absorption spectrum of monolayer $MoS_2$ at gate voltage of -100 V (Fig. 1d) corresponds to that of a nearly undoped sample. It shows pronounced absorption peaks from excitonic transitions, rather than the steps that would be expected for absorption arising from band-to-band transitions in 2D. The two features, known as the A (~ 1.92 eV) and B (~ 2.08 eV) excitons[2,3,20], are associated with direct optical transitions from the highest spin-split valence bands to the lowest conduction bands. These transitions are significantly modified by strong Coulomb interactions between the photogenerated electron-hole pairs and the corresponding formation of bound excitons.

In our investigation of the optical response of $MoS_2$ in the presence of a 2D electron gas (2DEG), we focus on the behavior of the low-energy (A) exciton over the spectral range of 1.8 - 2.0 eV. Figure 2a (red lines) shows representative absorption spectra of a monolayer $MoS_2$ FET under gate voltages from -100 to 80 V, corresponding to approximate charge neutrality to doping density of ~ $10^{13}$ cm$^{-2}$. We see an overall suppression of the optical absorbance in this spectral region with increasing electron doping. The prominent A exciton peak evolves into *two resonances,* with the emergence of a lower energy resonance (labeled as A$^-$). For $V_g > 0$, the absorbance of feature A diminishes rapidly and disappears into the background. The A$^-$ feature, on the other hand, broadens gradually, while approximately preserving its spectral weight. Similarly, in the PL spectra (Fig. 2a), both resonances can be identified for negative gate voltages. The PL intensity of the A exciton, like its absorbance, can be switched off by doping. In addition, redshifts in the PL peak energies from the corresponding absorption energies (Stokes shifts) are observed for both features; the magnitude of Stokes shift is found to increase with doping level (Fig. S1b).

What is the origin of the A$^-$ resonance? Our observation cannot be explained by a simple state-blocking effect. Although such Pauli blocking can account for the overall reduction in absorption with doping[21], it cannot explain the appearance of two resonances with different doping dependences. Nor can the emergence of two features be explained by an electric-field-induced band structure modification. In the absence of magnetic fields, the VS degeneracy of the bands is protected and no available degeneracy can be lifted to give rise to two distinct resonances. The observed A⁻ feature, however, can be understood arising from trions. Such bound states of two electrons to a hole, *i.e.,* negatively charged excitons, are known to have finite binding energies[10,11,22,23]. In our experiment at low temperature, the optical response of monolayer MoS$_2$ is dominated by neutral excitons in undoped samples. Trions emerge, accompanied by a reduction of exciton absorption and PL, when excess electrons are introduced to bind to photoexcited electron-hole pairs. The exciton spectral weight is transferred to the trion, as shown in our experiment, and to the low-frequency Drude response, as represented by the increased conductance of our sample. The detailed dependence on doping density of this phenomenon in the relatively low doping regime is often governed by electron localization in inhomogeneous samples.

We have analyzed the absorption spectra to determine the energies to create both neutral excitons ($\omega_A$) and trions ($\omega_A^-$) as a function of the doping level. The analysis is based on a fitting procedure using the predicted power-law spectral dependence[21,24] of the features (see Methods for details). We also convert the gate voltage to the electron doping density ($ne = CV_g$) using the back-gate capacitance of $C = 1.2 \times 10^{-8}$ Fcm$^{-2}$ and then to the Fermi energy ($E_F = \hbar^2 \pi n / 2 m_e e^2$) using an electron band mass of $m_e = 0.35 m_0$,[14] where $m_0$ is the electron mass. Because of intrinsic charging effects, we furthermore introduce an offset in the gate voltage of −107±6 V for $E_F = 0$. This value was extrapolated from the gate dependence of the PL Stokes shift[21,25] (Supplementary Information S2).

When the Fermi energy is varied from ~ 0 to 30 meV, the exciton energy blue shifts monotonically, while the trion energy remains largely unchanged, after a slight initial redshift (Fig. 2b). These dependences arise from the combined effects of Pauli blocking and many-body interactions[25,26], and are consistent with the previous theoretical results (Supplementary Information S3) and experimental studies on QWs[10-13]. The splitting between the exciton and trion energy is predicted to be linearly dependent on the Fermi energy and obey[11]

$$\omega_A - \omega_A^- = E_{A^-} + E_F, \qquad (1)$$

where $E_{A^-}$ is the trion binding energy. Since the exciton can be considered as an ionized trion, $\omega_A - \omega_A^-$ defines the minimum energy for the removal of one electron from the trion. In the limit of infinitesimal doping, it is just the trion binding energy $E_{A^-}$, the energy required to promote one of the electrons in a trion to the conduction band edge. At finite doping densities, an exciton is obtained by dissociation of a trion and placing the extra electron at the Fermi level, all lower conduction band levels being occupied at zero temperature (inset of Fig. 2c). Our experimental result for $\omega_A - \omega_A^-$ agrees very well with Eq. (1) (Fig. 2c), further confirming the spectroscopic assignment of the exciton and trion features. We also determine the trion binding energy from the linear fit to be 18.0±1.5 meV. This value is compatible with an estimated trion binding energy of $E_{A^-} \sim 0.1\ E_A$ (for isotropic 2D semiconductors with equal electron and hole masses[23]) and the calculated exciton binding energy of $E_A \approx 0.6$ eV[14].

The large trion binding energy observed in monolayer MoS$_2$ suggests the importance of trions even at elevated temperatures. Indeed, both the exciton and trion features can be identified in the optical response of doped monolayer MoS$_2$ at room temperature (Fig. 3a, b), although the resonances are significantly broadened. The absorption and especially PL are highly doping dependent (Fig. 3c). While the trion PL is largely gate independent, the exciton PL varies by two orders of magnitude, which is also correlated with the $I_{ds}$-$V_g$ dependence[27]. This dependence arises primarily from a spectral weight reduction of the exciton resonance with doping, which is consistent with the low temperature data (although less noticeable at low doping levels due to the finite temperature effects). The tuning range of the exciton PL could be further increased with better quality samples. Such strong tunability of the exciton PL was not observed at 10 K, where emission originates from hot PL before the exciton-trion equilibrium is reached, but at room temperature, exciton PL originates primarily from populations thermally excited from the trion state.

Finally we demonstrate the unique spin and valley properties of trions in monolayer $MoS_2$.[7] Fig. 4a shows the PL spectrum for a monolayer $MoS_2$ sample on a hexagonal boron-nitride (h-BN) substrate at 10 K, where the trion and exciton features are fully resolved because of the high sample homogeneity. We optically pump the sample at 1.96 eV, nearly on-resonance with the A exciton, by left circularly polarized light ($\sigma_-$). The trion emission is found to be nearly 100% of the same helicity as the pump light (Fig. 4b). The selection rules dictate the creation of an *e-h* pair in the K-valley with hole spin up and electron spin down upon absorption of a $\sigma_-$ photon. Our PL result indicates that the optically excited hole in the trion remains spin up, *i.e.,* in the K-valley (for a time much longer than the trion lifetime). The existence of tightly bound trions with dynamically controllable hole valley and spin in monolayer $MoS_2$ opens up possibilities of novel many-body phenomena. The large trion binding energy in 2D atomic crystals will also have implications for efficient manipulation of optical excitation by a bias electric field for novel optoelectronic and energy harvesting concepts.

**Methods**

**Sample preparation:** Monolayer $MoS_2$ samples were mechanically exfoliated from bulk $MoS_2$ crystals (SPI) on silicon substrates covered by a 280 nm layer of thermal oxide. Monolayer samples, typically of an area > 50 μm$^2$, were identified by optical microscopy. The sample thickness was confirmed independently by atomic-force microscopy, Raman and PL measurements[2,28]. Monolayer $MoS_2$ FET devices were fabricated by defining drain source contacts either through e-beam lithography or photolithography, followed by evaporation of Cr/Au or Ni in an e-beam evaporator. For the polarization resolved measurement, the exfoliated samples were transferred to hexagonal boron-nitride substrates.

**Optical measurements:** Monolayer $MoS_2$ FETs were measured at 10 K and at room temperature when the back-gate was varied systematically from -100 to 80 V. No significant hysteresis was observed. The drain-source current $I_{ds}$ was measured in the linear response regime under a drain-source bias voltage $V_{ds}$ = 40 mV across a channel length of 10 μm. For optical absorption measurements, the broadband radiation from a quartz tungsten halogen source was focused onto monolayer $MoS_2$ by a 40 × objective. The reflected radiation was collected and analyzed with a grating spectrometer equipped with a liquid nitrogen cooled CCD.

The reflectance contrast was measured by normalizing the radiation reflected from the sample on substrate to that from the bare substrate. The sample absorbance was obtained from the reflectance by using a Kramers-Kronig constrained variational analysis[29] (Supplementary Information S4). PL measurements were performed using the same setup with a single-mode solid-state laser centered at 532 nm. The laser intensity on the sample was kept below 500 Wcm$^{-2}$, which corresponds to a steady-state photoexcitation density of ~$10^{10}$ cm$^{-2}$ in monolayer MoS$_2$[2,30]. This is ~ 2 – 3 orders of magnitude lower than the electron doping density induced by the silicon back-gate. For the polarization resolved measurements, a Babinet-Soleil compensator was used to produce circularly polarized 632.8 nm light from a HeNe laser. Emission from the sample passed through a combination of a quarter-wave Fresnel rhomb and an analyzer for selection of the left and right circularly polarized components.

**Analysis of the absorption spectra:** We used the predicted single-sided power-law spectral dependence of the exciton and trion absorbance in doped 2D semiconductors, $A(\omega) = \sum_{X=A,A^-} a_X(\omega - \omega_X)^{-\alpha_X}$ (for $\omega > \omega_X$)[21,24]. This expression corresponds to a shakeup process, a non-instantaneous evolution of the 2DEG to a new equilibrium configuration upon creation of a hole[21,24] Here $a_X$, $\omega_X$ and $\alpha_X$ are, respectively, the amplitude, threshold energy and critical exponent of resonance $X$ (= A and A$^-$). In the limit of relatively low doping densities, as in our experiment, $a_X$ ($X$ = A, A$^-$) are weakly dependent on doping[24] and were kept as constants. To compare with experiment, the power-law dependence was convoluted with a Gaussian to account for sample inhomogeneity and the onset of indirect optical transitions in doped semiconductors[21]. In addition, a smooth background describing the low-energy tail of higher energy transitions was subtracted from experimental spectra. (See Supplementary Information S3 and S4 for details.) The fits (green lines, Fig. 2a), with individual contributions from exciton and trion (blue lines), are in good agreement with experiment (red lines). At positive gate voltages the exciton absorption is very weak and is omitted from the fitting.

**Acknowledgments** This research was supported by the National Science Foundation through grants DMR-0907477 (at Case Western Reserve University), DMR-1106172 and DMR-1122594 (at Columbia University).

**Author Contributions** K.F.M. and J.S. designed the experiment, performed the measurement and analysis, and prepared the manuscript; K.H. fabricated MoS$_2$ FET devices and measured PL; C.L. and J.H. developed MoS$_2$ FET devices; G.H.L fabricated MoS$_2$ samples on BN; T.F.H. contributed to the interpretation of the results and writing of the manuscript.

**Figures**

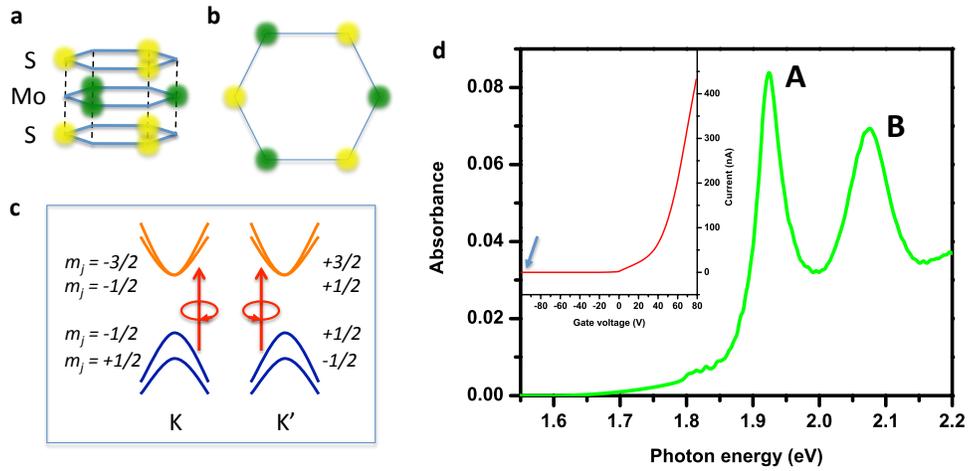

**Figure 1. Atomic structure, electronic band structure, and absorption spectrum of monolayer MoS$_2$.** **(a)** Representation of the trigonal prismatic structure of monolayer MoS$_2$. **(b)** Honeycomb lattice structure with each sublattice occupied by a Mo and two S atoms. **(c)** The lowest-energy conduction bands and the highest-energy valence bands labeled by the *z*-component of their total angular momentum near the K and K' point of the Brillouin zone. The spin degeneracy at the valence-band edges is lifted by the spin–orbit interactions. The valley and spin degrees of freedom are coupled. Under left circularly polarized excitation, only the K-valley is populated, while under right circularly polarized excitation, the K'-valley. **(d)** Absorption spectrum of undoped monolayer MoS$_2$ with two prominent resonances, known as the A and B excitons. The inset shows the drain-source current $I_{ds}$, under a $V_{ds}$ = 40 mV bias voltage, as a function of back-gate voltage $V_g$. It is characteristic of an *n*-doped semiconductor that can be turned off at large negative gate voltages.

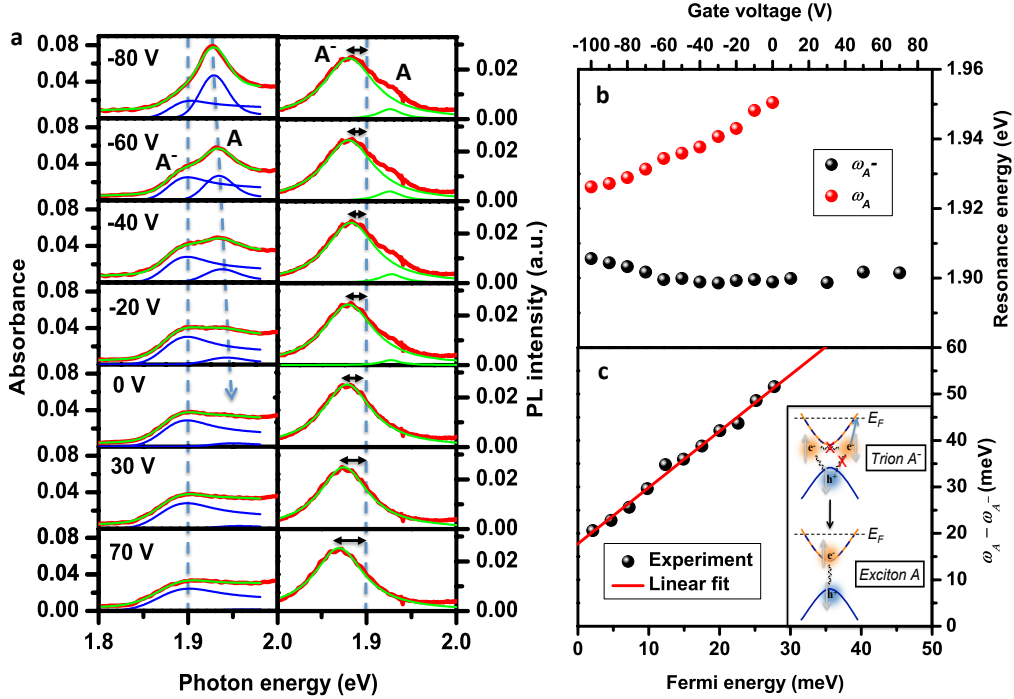

**Figure 2. Doping dependence of the optical properties of a monolayer MoS$_2$ FET. (a)** Absorption and PL spectra (red lines) in the range of 1.8 - 2.0 eV for the indicated back-gate voltages. The exciton (A) and trion (A$^-$) resonance behave differently with gate voltage. Left: Absorption spectra, with the dashed blue lines as a guide to the eye for the threshold energies of A and A$^-$ features. The green lines are power-law fits to the experimental results, as described in the main text, with the A and A$^-$ components shown as the blue lines. Right: The PL spectra of the A and A$^-$ features are fit to Lorentzians (green lines). The dashed blue line indicates the absorption peak of the A$^-$ resonance and the arrows show the doping dependent Stokes shift of the trion PL. **(b)** Threshold energies of the trion $\omega_{A^-}$ (black symbols) and the neutral exciton $\omega_A$ (red symbols), determined from the absorption spectra, as a function of gate voltage (upper axis) and Fermi energy $E_F$ (lower axis). **(c)** The difference in the exciton and trion energies, $\omega_A - \omega_{A^-}$ (symbols), as a function of Fermi energy $E_F$. The red line, a linear fit to the $E_F$-dependence, has a slope of 1.2 and an intercept of 18 meV. The latter determines the trion binding energy. Inset: representation of the dissociation of a trion into an exciton and an electron at the Fermi level.

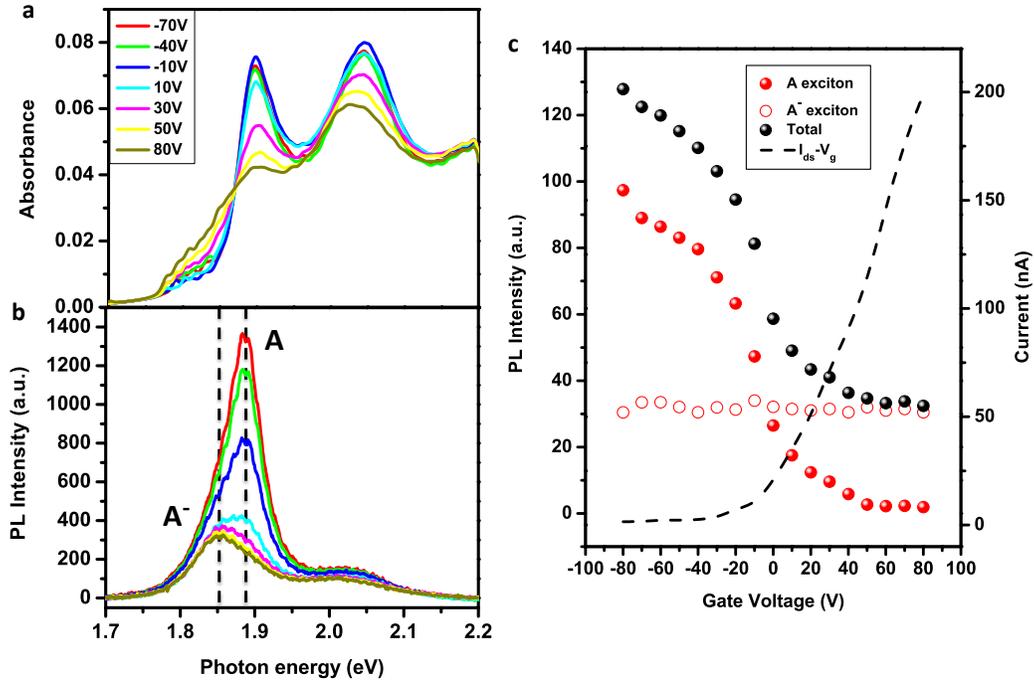

**Figure 3. Observation of excitons and trions at room temperature in monolayer MoS$_2$. (a,b)** Absorption and PL spectra at different back-gate voltages. Both the neutral exciton (A) and trion (A$^-$) features can be identified, although the resonances are significantly broadened. **(c)** Dependence on gate voltage of the drain-source current (right) and the integrated PL intensity of the A, A$^-$ features and their total contribution (left).

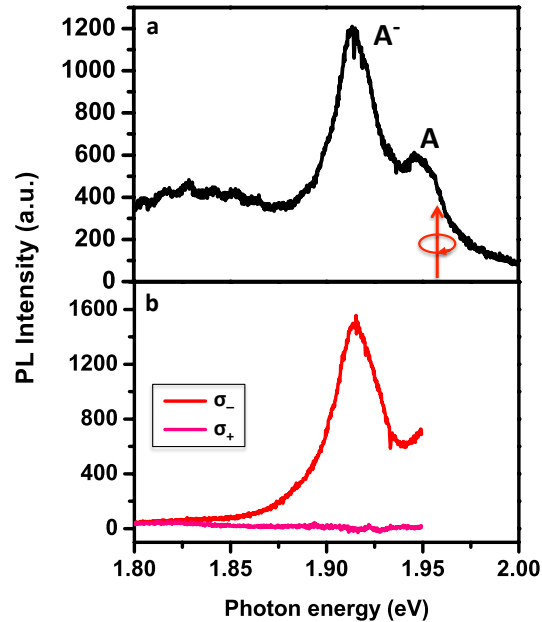

**Figure 4. Valley and spin control of trions in monolayer MoS$_2$. (a)** PL spectrum of a monolayer MoS$_2$ sample on a h-BN substrate at 10 K with fully resolved exciton (A) and trion (A$^-$) emission. **(b)** The left circularly polarized ($\sigma_-$) and right circularly polarized

($\sigma_+$) component of the trion emission for excitation by $\sigma_-$ light at 1.96 eV, nearly on-resonance with the A exciton.

# Supplementary Information
## Observation of tightly bound trions in monolayer $MoS_2$
Kin Fai Mak, Keliang He, Changgu Lee, Gwan Hyoung Lee, James Hone, Tony F. Heinz, and Jie Shan

## S1. Estimate of the interaction parameter $r_s$ in monolayer $MoS_2$

The electron interaction strength in a two-dimensional electron gas (2DEG) system is characterized by the dimensionless parameter $r_s$, which measures the average interparticle separation in units of the effective Bohr radius. The parameter can also be considered to be the ratio of the average Coulomb potential energy to the average kinetic energy for the 2DEG. For a given total electron density $n$, the $r_s$ is given by

$$r_s = \frac{g_v}{\sqrt{\pi n}(\varepsilon_{eff}\hbar^2/m_e e^2)}, \quad (S1)$$

where $m_e$ and $\varepsilon_{eff}$ are, respectively, the electron effective mass and the background dielectric constant, $g_v (= 2)$ is the valley degeneracy for $MoS_2$, $\hbar$ is Planck's constant, and $e$ is the elementary charge. Spin degeneracy (= 2) has been taken into account in Eq. S1. For an $n$-doped monolayer $MoS_2$, we use an electron band mass $m_e = 0.35 m_0$ (in electron mass $m_0$) from GW calculations of the electronic structure[1,2] and $\varepsilon_{eff} = \sqrt{\varepsilon_\parallel \varepsilon_\perp}$, with the in- and out-of-plane dielectric constants of $\varepsilon_\parallel = 2.5$ and $\varepsilon_\perp = 6.76$.[3] The former is the average of air and $SiO_2$ dielectric constants for our monolayer $MoS_2$ samples on $Si/SiO_2$ substrates. We then obtain $r_s \approx 60$ at a charge density of $n = 1 \times 10^{11}$ cm$^{-2}$. This value of $r_s$ is almost an order of magnitude higher than that of 2DEG systems in conventional semiconductor quantum wells at the same electron density[4]. Such a large interaction parameter, achievable in the presence of relatively high carrier densities, indicates that monolayer $MoS_2$ should be an ideal system for studies of many-body phenomena, such as carrier multiplication and Wigner crystallization.

## S2. Calibration of the Fermi energy in monolayer $MoS_2$ based on the doping dependence of the exciton photoluminescence Stokes shift

As discussed in the main text, Stokes shifts were observed for both the exciton and trion photoluminescence (PL) in our monolayer $MoS_2$ field-effect transistors (FETs) and these shifts were dependent on the gate voltage. Here we discuss how the gating voltage corresponding to the Fermi energy $E_F$ at the bottom of the conduction band (defined as $E_F = 0$) was determined. We note that because of the presence of trapped charges, this condition does not correspond to zero gate voltage.

The gate dependence of the exciton and trion energies, determined from the absorption and PL spectra, is shown in Fig. S1(a). The threshold energies for absorption were extracted from the absorption spectra using single-sided power-law fits (see Section S3 below for details). The peak energies for PL were extracted from the PL spectra using multiple Lorentzian fits. Fig. S1(b) shows the Stokes shift of the exciton and trion PL. The Stokes shift of both the exciton and of the trion (outside the very low doping regime) is seen to vary linearly with gate voltage. We extrapolate the linear dependence to obtain a gate voltage of −107±6 V for which $E_F = 0$.

To understand the doping dependence of the PL Stokes shift and the above calibration procedure, let us first consider the absorption and PL process in a doped direct-gap semiconductor without the many-body effects. In accordance with our experiments, we assume a degenerate electron gas. As illustrated in Fig. S1(c), the threshold energy for absorption is given by $\omega_D$, corresponding to the onset of direct dipole transitions[5]. The absorption, however, extends to a lower energy, $\omega_I$, because of the presence of indirect optical transitions. These indirect transitions broaden the threshold absorption feature to the red by a width corresponding to $E_F$. On the other hand, the luminescence process (excluding hot luminescence) takes the system from the lowest-energy excited state back to the ground state (with emission at energy $\omega_I$), as well as to final states involving excitations of the 2DEG. The PL spectrum thus extends from the band-gap energy up to $\omega_I$ with a broadening also governed by $E_F$[5-8]. Therefore, the separation between the absorption and PL peak energy, *i.e.* the Stokes shift, is proportional to $E_F$[5-7], which in turn is linear with the gate voltage because of the constant density of states for massive particles in 2D.

The above conclusion does not change when the many-body effects are taken into account[5]. We also note that while the exciton PL Stokes shift follows the predicted linear dependence on doping, deviation from the linear doping dependence was observed for the trion PL Stokes shift in the low doping regime. This is likely due to the presence in our samples of potential fluctuations, such as defects or impurities. At low doping densities, the dielectric screening is much reduced and the trions could be significantly influenced by the potential fluctuations[9,10]. The excitons being neutral particles, on the other hand, are expected to be less sensitive to the potential fluctuations.

**S3. Many-body optical response of a doped 2D semiconductor**

The optical response of a doped direct-gap semiconductor in 2D has been extensively investigated theoretically[5,8,11,12]. At low-temperature, the many-body interactions are predicted to give rise to a single-sided power-law dependence for the absorption spectrum $A(\omega)$ at photon energy $\omega$[5,8,11,12]:
$$A(\omega) = \sum_{X=A,A^-} a_X (\omega - \omega_X)^{-\alpha_X}, \qquad \omega > \omega_X. \tag{S2}$$
Here $\omega_X$, $a_X$ and $\alpha_X$ are, respectively, the absorption threshold energy, amplitude and exponent for the relevant excitation $X = (A, A^-)$. Two types of excitations (excitons and trions) have been observed in our experiment and are included in Eq. S2. In general there could be other types of excitations, such as continua. The power-law spectral dependence arises from a real-time dependence of $t^{-1-\alpha_X}$ for the optical conductivity of the *n*-doped 2D semiconductor following the creation of a hole by optical excitation[5,8]. It reflects the evolution of the 2DEG to a new equilibrium configuration of excitation $X$[5,8].

We note that the amplitude $a_X$ is expected to depend only weakly on doping level in the regime of relatively low densities in our experiment (where the Fermi wavelength significantly exceeds the exciton and trion Bohr radii). This has been verified for the explicit expressions for $a_X$ in Ref. 8. For simplicity, we assume the amplitudes to be independent of carrier density in our analysis.

The threshold energies for the trion and exciton absorption are given as[5,6,8,13]
$$\omega_{A^-} = E_0 + E_\delta - E_{A^-} - E_A,$$
$$\omega_A = E_0 + E_\delta + E_F - E_A. \tag{S3}$$
Here $E_0 = E_g + \hbar^2 \pi n / 2\mu e^2$ (with the renormalized band-gap $E_g$ and the *e-h* reduced mass $\mu$: $1/\mu = 1/m_e + 1/m_h$), $E_\delta$ is the ground state energy shift of the 2DEG due to the presence of the hole, and $E_{A^-}$ and $E_A$ are, respectively, the trion and exciton binding energy. The origin of the threshold energies can be understood from the schematic representation of Fig. S2. Energy $E_0$ is required to generate a free *e-h* pair in the doped semiconductor without many-body effects. Upon the optical generation of a hole, the 2DEG responds to it, and the quasi-particle states acquire a shift in energy $E_\delta$. This shift can be evaluated from the electron scattering phase shift $\delta_{\mu\nu}(\varepsilon)$ by summing over all energies $\varepsilon$, spins $\mu$ and valleys $\nu$:[14] $E_\delta = -\int_0^{E_F} d\varepsilon \sum_{\substack{\mu=\uparrow,\downarrow \\ \nu=1,2}} \frac{\delta_{\mu\nu}(\varepsilon)}{\pi}$. We note that the energy shift is negative and lowers the excitation threshold energies. Now to form a trion, two electrons from the 2DEG are bound to the hole with a total binding energy of $E_{A^-} + E_A$. We obtain the trion threshold energy given in Eq. S3. Similarly, to form a neutral exciton, one of the two electrons in the trion is unbound ($-E_{A^-}$) and placed at the Fermi surface of the 2DEG ($+E_F$).

The exponents $\alpha_X$ are determined by the number of electrons $n_{\mu\nu,X}$ induced by the optically created hole potential for its equilibrium configuration $X = (A, A^-)$ [5,13,15]: $\alpha_X = 1 - \sum_{\substack{\mu=\uparrow,\downarrow \\ \nu=1,2}} n_{\mu\nu,X}^2$. In a simple model of a spin singlet exciton and a trion comprised of two electrons of opposite spin, the exponents can be related to the scattering phase shift at $E_F$ by Hopfield's rule of thumb[5,8,13,15]:

$$\alpha_{A^-} = 1 - \sum_{\nu=1,2}\left[\frac{\delta_{\downarrow\nu}(E_F)}{\pi} - 1\right]^2 - \sum_{\nu=1,2}\left[\frac{\delta_{\uparrow\nu}(E_F)}{\pi}\right]^2 \quad \text{(S4)}$$

$$\alpha_A = 1 - \sum_{\mu=\uparrow,\downarrow;\nu=1,2}\left[\frac{\delta_{\mu\nu}(E_F)}{\pi} - 1\right]^2.$$

The difference between the exponents for the exciton and the trion is a consequence of the differing number of electrons in the 2DEG that can screen out the hole potential in the configurations of exciton and trion.

We use Eq. S2 to analyze the experimental absorption spectra for the spectral range of A and B exciton at different doping levels. First, a small background accounting for the tails of higher energy transitions was subtracted. This contribution, as discussed in section S4 below, was determined independently from experimental results for bulk $MoS_2$ and was assumed to be independent of doping and sample thickness. Such assumptions are justified by the relatively weak doping in our experiment and the fact that the higher energy transitions are at least 1 eV above those being analyzed. Furthermore, earlier experimental studies of few-layer $MoS_2$ up to 6 layers have shown a weak dependence of higher energy transitions on the sample thickness[16]. Eq. S2 was then convoluted with a Gaussian to account for sample inhomogeneity and the onset of indirect optical transitions, as discussed above in Section 2. The fits of Eq. S2 to the absorption spectra, together with the trion and exciton components, are shown in Fig. 2a of the main text. Good agreement has been obtained for spectra at all doping levels.

The fitting parameters for the exciton and trion threshold energies are presented in Fig. 2b of the main text. Although each varies differently with doping level, the energy separation $\omega_A$ - $\omega_{A^-}$ depends linearly on $E_F$, precisely as predicted by Eq. S3. This behavior verifies our assignment of the states and permits an accurate determination of the trion binding energy $E_{A^-}$. The fitting parameters for the exponents are shown in Fig. S3. For nearly intrinsic samples, $\alpha_{A^-} \approx 0$ and $\alpha_A \approx 1$. These parameters are consistent with the fact that the exciton resonance is prominent, while the trion resonance is completely absent. With increasing doping, the exponent for the exciton $\alpha_A$ decreases, corresponding to the experimental observation of the reduction of the exciton resonance. On the other hand, the trion exponent increases with doping and then flattens out after reaching certain doping densities. This corresponds to the observation of the emergence of the trion with doping and its independence on doping beyond a certain level. The power-law exponents of Fig. S3 for exciton and trion absorption are also compatible with Eq. S4 for the expected variation of the electron scattering phase shifts, which decreases from π to 0 with increasing doping.

**S4. Conversion of the reflectance spectrum to the absorption spectrum of monolayer MoS$_2$ by a Kramers-Kronig constrained variational analysis**

We determine the absorbance spectrum $A(\omega)$ of monolayer MoS$_2$ from the experimental reflectance spectrum at normal incident for monolayer MoS$_2$ on Si/SiO$_2$ substrates. A Kramers-Kronig constrained analysis[17] was applied because simultaneous transmission measurements were not possible for our structures.

The optical reflectance from MoS$_2$ samples on the substrate, $R$, and the reflectance from the bare substrate, $R_0$, can be calculated from a multilayer analysis as[18]

$$R(\omega) = \left| -\frac{Cn_{SiO_2} - D[1-A(\omega)]}{Cn_{SiO_2} + D[1+A(\omega)]} \right|^2, \quad R_0(\omega) = \left| -\frac{Cn_{SiO_2} - D}{Cn_{SiO_2} + D} \right|^2. \tag{S5}$$

The normalized reflectance is given by $R/R_0$. Here the coefficients $C$ and $D$ are related to the speed of light in vacuum $c$, the refractive index $n_\zeta$, thickness $L_\zeta$ and phase shift $\delta_\zeta = n_\zeta \omega L_\zeta / c$ in the $\zeta$ = SiO$_2$ and Si layers:

$$\begin{pmatrix} C \\ D \end{pmatrix} = \begin{pmatrix} n_{SiO_2} \sin\delta_{SiO_2} & in_{Si}\cos\delta_{SiO_2} \\ in_{SiO_2}\cos\delta_{SiO_2} & n_{Si}\sin\delta_{SiO_2} \end{pmatrix} \begin{pmatrix} \sin\delta_{Si} + in_{Si}\cos\delta_{Si} \\ i\cos\delta_{Si} + n_{Si}\sin\delta_{Si} \end{pmatrix}. \tag{S6}$$

The absorbance $A(\omega)$ of monolayer MoS$_2$ can be expressed through the complex dielectric function $\varepsilon(\omega)$ of the monolayer and its thickness $d$ (taken as $d$ = 0.67 nm, corresponding to the bulk layer spacing) as $A(\omega) = \frac{-i\omega d}{c}\varepsilon(\omega)$. Note that the real part of the complex quantity $A(\omega)$ corresponds to the absorption that would be expected for transmission through the MoS$_2$ monolayer in vacuum.

We model the dielectric function of the MoS$_2$ monolayer by a sum of Lorentzians, so that the Kramers-Kronig relations are automatically satisfied. For each gate voltage, we vary the parameters of the Lorentzians so that Eqs. S5 and S6 provide the best fit to the experimental normalized reflectance. Fig. S4a shows the experimental normalized reflectance spectra (red curves) and the fits (green curves); Figs. S4b and S4c are the corresponding real and imaginary part of the absorbance spectra based on the inferred dielectric function.

In our implementation, we used 200 equally spaced Lorentzians to describe the dielectric response of monolayer MoS$_2$ over the spectral range of 1.7 - 2.2 eV, covering the A (~ 1.92 eV) and B (~ 2.08 eV) excitons:

$$\varepsilon(\omega) = \varepsilon_{con}(\omega) + \sum_j \frac{\omega_{p,j}^2}{\omega_j^2 - \omega^2 + i\Gamma\omega}. \tag{S7}$$

Here $\omega_{p,j}$ and $\omega_j = 1.7 + j \times 0.0025$ (in eV) are the plasma and resonance energy of the $j$-th oscillator (Lorentzian). The broadening parameter $\Gamma$ was fixed at 10 meV, a few times larger than the spacing between the oscillator resonance energies. In Eq. S7 we also included a term $\varepsilon_{con}(\omega)$ to account for contributions from oscillators above 2.2 eV. We estimated $\varepsilon_{con}(\omega)$ by fitting the complex dielectric function from an earlier experiment for bulk MoS$_2$ over a very broad spectral range[19] using Lorenzians only. The difference in the spectral range of 1.7 - 2.2 eV between the experimental value and the sum of Lorenzians peaked below 2.2 eV was assigned to $\varepsilon_{con}(\omega)$ and modeled by a polynomial with $\omega$ in eV

$$\varepsilon_{con}(\omega) = -46.20 + 57.96\omega - 11.40\omega^2 + i(56.28 - 71.04\omega + 22.40\omega^2). \quad \text{(S8)}$$

We assume in this analysis that $\varepsilon_{con}(\omega)$ is independent of layer thickness and doping.

To fit the experimental normalized reflectance spectrum at a given gate voltage, the plasma energies $\omega_{p,j}$ ($j$ = 1 to 200) were used as free parameters. We used the known layer thicknesses $L_{SiO_2}$ = 280 nm, $L_{Si}$ = 0.5 mm, and the dispersion of $n_{SiO_2}$ and $n_{Si}$ from Ref. 20. We found that 200 Lorentzians were adequate to fit the relatively smooth reflectance spectra.

**Supplementary figures**

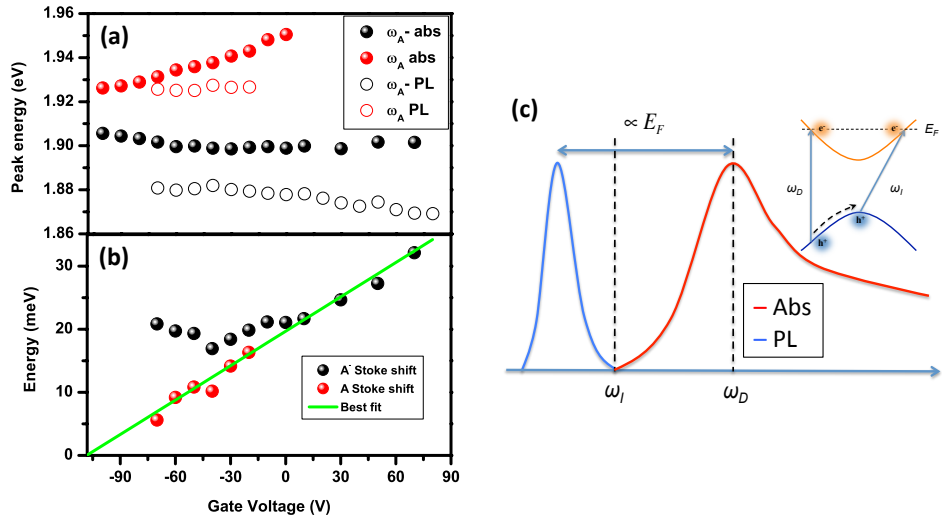

Figure S1. (a) Measured dependence of the absorption threshold energy and the peak PL energy of the trion (black) and exciton (red) on gate voltage. (b) Gate dependence of the Stokes shift for the exciton (A, red) and trion (A⁻, black) PL. The green line is a linear fit to the gate dependence of the Stokes shift of the exciton PL and trion PL outside the low doping regime. (c) Schematic representation of the absorption and PL process in a doped direct-gap semiconductor, including the onset of indirect optical transitions at energy $\omega_I$ and direct optical transitions at energy $\omega_D$ (many-body effects not considered). The PL Stokes shift is proportional to the Fermi energy.

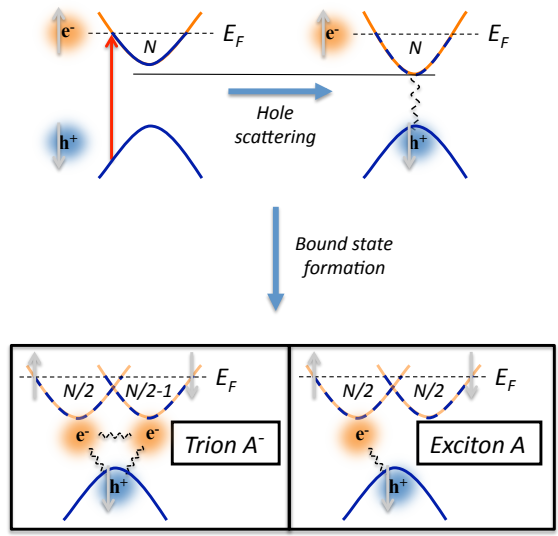

Figure S2. Schematic representation of the optical excitation of a trion and exciton in a doped direct-gap semiconductor in 2D. An *e-h* pair is created by absorbing a photon across the fundamental energy gap. With the introduction of the hole potential, the N electrons in the 2DEG scatter off the hole potential and lower the energy of the quasi-particle states. The N+1 electron system then shares one electron of the opposite spin as of the hole to form a singlet exciton A (right) or shares two electrons to bound to the hole to form a trion $A^-$ (left). Here two electrons of opposite spins are shown as an example.

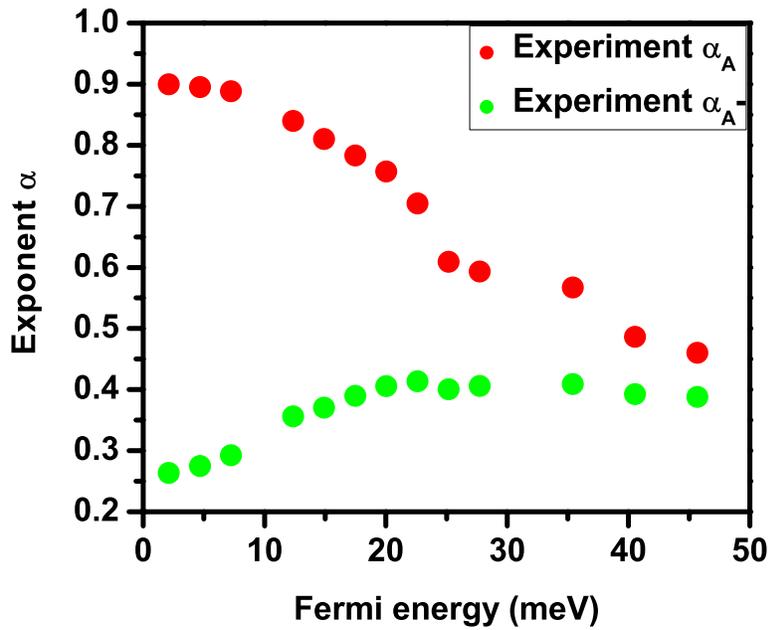

Figure S3. The absorption power-law exponents of Eq. S2, inferred from fitting the experimental absorption spectra for the exciton (red) and trion (green), as a function of the Fermi energy.

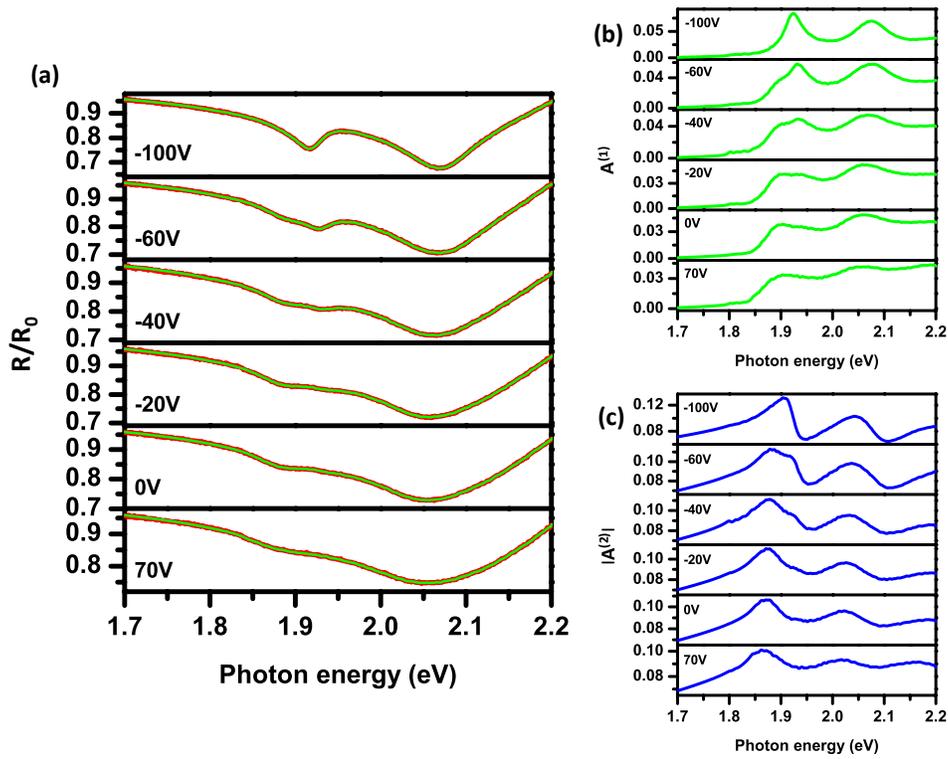

Figure S4. (a) Reflectance spectra of monolayer $MoS_2$ on a $Si/SiO_2$ substrate normalized by reflectance spectra of the substrate. The different panels correspond to the indicated gate voltages. The red and green curves are, respectively, the experimental spectra and the fits from the Kramers-Kronig constrained analysis described in the text. (b) and (c) are the real and imaginary parts of the optical absorbance spectra for monolayer $MoS_2$ deduced from this procedure.

**Supplementary references**